\documentclass[aps,prl,twocolumn,superscriptaddress,nofootinbib,floatfix,
nobalancelastpage,nobibnotes]{revtex4}
\usepackage{epsfig}
\usepackage{graphicx}
\usepackage{caption}

\begin{document}

\title
{Neutrino collective effects during their decoupling era in the early universe}

\author{R. F. Sawyer}
\affiliation{Department of Physics, University of California at
Santa Barbara, Santa Barbara, California 93106}

\begin{abstract}
There is an accepted approach to calculation of the neutrino flavor density-matrix in the halo of a supernova, in which neutrino amplitudes, not cross-sections, need to be followed carefully in the region
above the region of frequent scatterings.
The same reasoning and techniques, applied
 to the evolution of neutrino flavors and energy distributions in the early universe in the era of neutrino decoupling, leads to radical changes in the predictions of the effects  of the neutrino-neutrino interaction.
 Predictions for the production of sterile neutrinos, should they exist, will also be changed.
\end{abstract}
\maketitle

\section{1. Introduction}
There has been an explosion of recent works \cite{jap} -\cite{ful} with the aim of checking and improving the results of the
standard calculations \cite{dol} of the neutrino distributions in the big bang during the era of neutrino decoupling. One of the attributes of this body of work is that it continues the practice of using neutrino-neutrino cross-sections in the temperature region around T=1 MeV, instead of following amplitudes in detail. This is in contrast to ``refractive effects",  meaning as calculated from forward amplitudes, as are indices of refraction, but here much complicated by many-body effects that dominate at very high number densities. 

All the experience from supernova halo calculations says we should be following the amplitudes. In the early universe the coherent amplitude approach is even more necessary, in our view of how the technology works. The basic equations for following amplitudes originate from the formalism of  \cite{rs}, and we follow them here, up to a point. Finally, in this paper we show that ``fast" collective processes \cite{rfs1}-\cite{f9} will dominate. 

We begin our investigations by looking at a beam of $\bar \nu_e$'s moving upward colliding with a beam of $ \nu_e$'s that moves downward, maximally violating the initial isotropy and near flavor balance that we shall demand later in this paper. 
Here, and in everything that follows, by ``beams" we mean streams that are very nearly, but not quite, unidirectional. We would like to use all three active $\nu$'s, but because of computational complexities we are limited to two. For now $\nu_x$ is either $\nu_\mu$ or $\nu_\tau$. We maintain equal numbers of neutrinos and anti-neutrinos at all times. Later we address spherically symmetric initial conditions.

The present work will be applied roughly to the temperature range
.1 MeV$ <  T <$10 MeV  and over time scales such that the 
$\nu$ masses can be neglected, both in their capacity as flavor mixers, and in the disruption of the coherence that is at the center of our methods. This time scale will be also many of orders of magnitude less than the mean time for ordinary scattering.
 For the initial neutrino states we take Fermi distributions in energy with chemical potentials equal to zero.
Now consider a reaction,
\begin{eqnarray}
\nu_e({\bf p})+  \bar \nu_e({\bf q})\leftrightarrow \nu_x({\bf p})+\bar \nu_x({\bf q})~  ,
\label{unstable1}
\end{eqnarray}
with momentum ``preservation", not just momentum conservation, but where flavors can be exchanged 
between the beams and the designations as $\nu$, or $\bar \nu$ can be exchanged.
This instability in this amplitude was first identified in refs. \cite{fchir},  \cite{raff3} and it produces a fast mode exchange of particle and anti-particle identity in a two-beam simulation. 
It cares not at all about individual particle energies, as long as all particles are sufficiently relativistic. 

The instability is robust, in a sense that early ``fast" process models , e.g. those of the present author \cite{rfs3x}, were not. In these latter models nature had to provide very specific forms of angular distributions for the instability to appear.  The instabilities that we shall discuss thrive on angular distributions ranging, we believe, from two beams head-on to isotropic.

\section{2. The model}

We consider only time intervals that are short on the scale of free paths for scattering, and in which the neutrino-neutrino interactions, if unstable, can dominate the dynamics. And in calculating coherent effects we must
trace what each quantum state of the multi-particle system does, before adding up to see how the statistical ensemble (which is where the isotropy, e.g., is encoded) evolves.

Let $\sigma_+^{ \bf{p}}$ be the operator that takes a $\nu_e$ of momentum ${\bf p}$ into a $\nu_x$ of the with $\sigma_3^{\bf{ p}} $ measuring the flavor in the usual way. Then the standard model interaction between this mode and one of momentum ${\bf q}$ is of the form,
\begin{eqnarray}
const. \times  [\sigma_+^{ \bf p} \sigma_-^{\bf q}+\sigma_-^{ \bf{p}} \sigma_+^{ \bf{q}}
+2^{-1}\sigma_3^{ \bf{p}} \sigma_3^{ \bf{q}}]\,,
 \end{eqnarray}
Where $\sigma_+=(\sigma_1+i \sigma_2)/2$, etc. and $\bar \sigma$'s act in the anti-particle space.  
 The operators $ \sigma^{ \bf p}$ and $ \sigma^{ \bf q}$ separately obey Pauli matrix commutation rules among themselves, as do the $\bar \sigma$'s. Generalizing to a system containing many modes and introducing antiparticles, the effective interaction is given in \cite{rs} as,
  \begin{eqnarray}
&H_{\rm eff}= { 4 N_a \pi G_F \over  V} \sum_{j,k}^{N_a} g_{j,k} \Bigr [\sigma_+^j \sigma_-^k +\sigma_-^j\sigma_+^k 
+{1\over 2} \sigma_3^j \sigma_3^k  \,+
\nonumber\\
\,
\nonumber\\
&\bar \sigma_+^j \bar \sigma_- ^k+\bar\sigma_-^j \bar\sigma_+^k
+{1\over 2} \bar \sigma_3^j\bar\sigma_3^k -\bar \sigma_+^j  \sigma_-^k -\bar \sigma_-^j \sigma_+^k 
-{1\over 2} \bar \sigma_3^j \sigma_3^k
\nonumber\\
\,
\nonumber\\
&-\sigma_+^j \bar \sigma_-^k -\sigma_-^j \bar \sigma_+^k
- {1\over 2}  \sigma_3^j \bar \sigma_3^k \Bigr ] \,.
\label{hammy}
\end{eqnarray}
Here $g_{j,k} =(1-\cos \theta_{j,k})$ in the applications considered below, but the definition could be extended, for example, for three-flavor simulations. 
Only terms that embody ``momentum preservation", as defined above, are to be included. 
 
 In the definition above, the indices $j,k$ initially served to label the various individual neutrinos.
 For our computations we grouped them into $N_a$ groups each equally occupied with $N/N_a$ neutrinos,
 each localized in a tiny region of solid angle, and each in an initial state of definite flavor.  Then we used the fact that,
 $\sum_{j,k}^{N_a} [\sigma_+^j, \sigma_-^k]=N_a\sum_k^{N_a} \sigma_3^k$ , etc.  to rescale, producing the $N_a$ factor in (\ref{hammy}).  Then the commutation rules for the $\vec \sigma $'s and $\vec \tau$'s are exactly those of the 2x2 case in our original description.  
 The $j$ and $k$ in (\ref{hammy}) now index the various groups, rather than the single neutrinos which occupy them. This is all possible only because neutrino energies do not enter $H_{\rm eff}$. A nearly mono-angular beam still contains a big range of values of neutron energy.

We refer to this as the effective Hamiltonian instead of $-\mathcal{L_I }$, because the particle number and kinetic
energies are taken as exactly conserved at every step of the process, so that the kinetic term in the Lagrangian is completely irrelevant. The dynamics is all in the flavors, and the wave functions are those in a periodic box of volume V  large enough to accommodate our phenomena.

We derive the equations of motion for the variables in (\ref{hammy}), using the commutation rules of Pauli matrices.
 Then we can rescale the time variable to eliminate the explicit coupling constant in the equations, with a time unit that
is $ ({ 4 \pi G_F N\over  V})^{-1}= ({ 4 \pi G_F n })^{-1}$, where $n$ is the number density. The Heisenberg equations  become,
\begin{eqnarray}
&i {d\over dt} \sigma_+^j=\sum _k^{N_a} g_{j,k} \Bigr [\sigma_3^j( \sigma_+^k  -\bar \sigma_+^k )+ \sigma_+^j(\bar \sigma_3^k -\sigma_3^k ) \Bigr ]~,
\nonumber\\
&i {d\over dt}\bar \sigma_+^j=\sum _k^{N_a} g_{j,k} \Bigr [\bar \sigma_3^j( \bar \sigma_+^k  - \sigma_+^k )+\bar\sigma_+^j( \bar\sigma_3^k -\sigma_3^k )\Bigr ]~,
\nonumber\\
&i {d\over dt} \sigma_3^j=\sum _k^{N_a} g_{j,k} \Bigr [2(\sigma_+^j \sigma_- ^k - \sigma_-^j \sigma_+^k )-
\nonumber\\
& 2(\sigma_+^j  \bar \sigma _- ^k - \sigma_-^j \bar\sigma_+^k)\Bigr ]~,
\,\nonumber\\
&i {d\over dt} \bar \sigma_3^j=\sum _k^{N_a} g_{j,k} \Bigr [-2(\bar\sigma_-^j\bar\sigma_+ ^k - \bar \sigma_+ ^j\bar\sigma_-^k )+
\nonumber\\
&2(\bar \sigma_-^j  \sigma_+^k  - \bar\sigma_+^j \sigma_-^k )\Bigr ]~.
\nonumber\\
&\,
\label{array}
\end{eqnarray}
These are still operator equations in principle. The ``mean field theory" (MFT) that we invoke is 
the assumption that we can now replace the operators in (\ref{array}) by their expectations in the medium.

First we search for the instabilities that characterize fast evolution
for the case $N_a=2$, an up-beam and a down beam. The coupling matrix $g$ in (\ref{array})
is then $g_{1,1}=g_{2,2}=0 ~; g_{1,2}=g_{2,1}=1$. The mixing terms 
$\sigma_ \pm^j, \bar \sigma_\pm ^j$   for (\ref{array}) all appear be zero at $t=0$, since normal $\nu$ mass terms do not provide any $\nu-\bar \nu$ mixing.
But we do know that in this case quantum effects can initiate evolution, and that they are closely simulated by taking initial values $\sigma_+^j= i N^{-1}$,
$\bar\sigma_+^j=- i N^{-1}$.  Our most recent and best discussion of this is in \cite{rfs7}  (see also \cite{rfs6})
but one of the progenitors is from a year 2000 paper on condensed matter issues \cite{va} and attributes the essential technique to Bogolyubov, in the ``BBKYG hierarchy" mode. It involves finding the leading correction to the mean-field theory's general assumption, $\langle A B \rangle=\langle A \rangle \langle B \rangle$, and then building a bigger algebraic structure before constructing an extended list of equations
in which we then replace operators by expectation values.

We firmly believe that these seeds, which stand in for actual very early times calculations, are not miscellaneous ``vacuum fluctuations", and are totally determined within the QFT defined by
$H_{\rm eff}$ of (\ref{hammy}).

Then looking at (\ref{array}) we see that the early change in the variables $\sigma^j_3,\bar \sigma^j_3$, 
is quadratic in the seeds, so we can take them as fixed at their initial values for a short time after turn-on, thus linearizing the equations for the $\sigma^j_{\pm}$ operators. 
We find an eigenstate of the response matrix corresponding to the initial state,
\begin{eqnarray}
[\sigma^1_3, \sigma^2_3, \bar \sigma^1_3, \bar \sigma^2_3] = [1,-1,1,-1] \,,
\end{eqnarray}
with imaginary eigenvalues $2^{-1/2} $ attached to the state,

$\nu_e({\bf p})+  \bar \nu_x({\bf q})+ \nu_x({\bf p})+\bar \nu_e({\bf q})$.

We show in fig. 1 results for the solutions of this example. The state at complete turnover, at around $t=4$, has returned to a very nearly flavor-diagonal mode with ``up" and ``down" essentially reversed. 

 \begin{figure}[h] 
 \centering
\includegraphics[width=2.5 in]{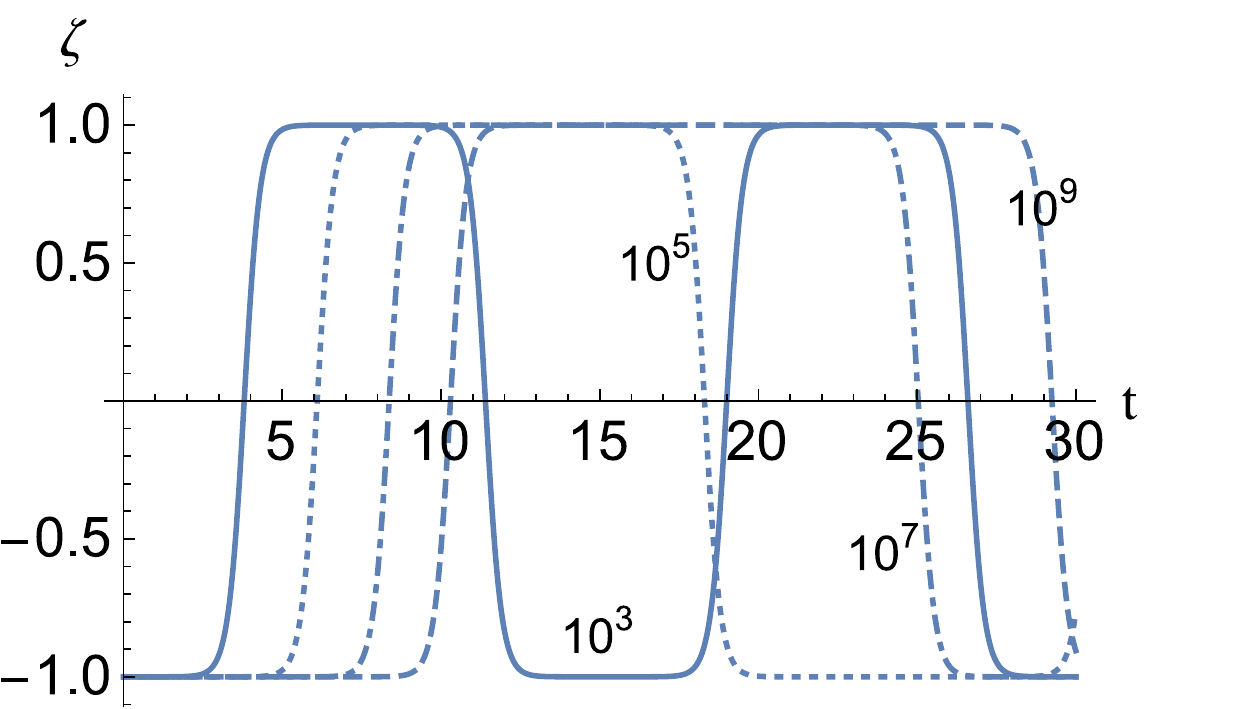}
 \caption{ \small } 
Solutions of (\ref{array})  $\zeta=\sigma_3=n_x-n_e$ potted against time, where the $n$'s are occupation averages in the up-moving beam that began as pure $\nu_e+\bar\nu_x$. For a number density (of each species) of $(3~{\rm MeV})^3$ the time scale is in units of $ \hbar \, 10^{5} ({\rm eV} )^{-1}$,  approximately. 
 N, ranging from $10^3$ to $10^9$,  is the number of particles in the simulation, which entered through the initial conditions for the 
 $\sigma_\pm$ operators
 \label{fig. 1}
\end{figure}
The equal spacings indicate an $N$ dependence of the turn-over time that is linear in $\log N$ for large $N$. 

\section{3. Multiple beams and the possum state}
In the preceding, our definition of single beam was a based on imagining a very tight bundle of $N$ directions, in an element 
of solid angle, $\delta \Omega <<4 \pi$, where each direction corresponded to that of a neutrino. The magnitudes of momenta will be drawn at random from a Fermi distribution with zero chemical potential at the temperature that prevails in the medium. 
In the two-beams-clashing calculation, and the plots in fig. 1, we can follow, say, the development
given an initial state that was initially
equal numbers of $\bar \nu_e$  and $\nu_x$ going up and of $\bar \nu_x$  and $\nu_e$ going down..

This manifestation of the instability at first inspection is not of much use in modeling a 1D universe,
whose first duty is to stay isotropic, in the up-down sense. So next we add another pair of up-and-down beams, this time taking the flavors reversed as compared those in the first beam. We started in a state of complete flavor equilibrium, in the mean-field sense, and stayed in equilibrium, so nothing should happen. That is confirmed in a MFT calculation that now has four up-moving beams and four down-moving ones. 

Then we had intended to test the effects of this background system, in which nothing is really happening on its own, on non-equilibrium perturbations such as the excess of $\nu_e-\bar\nu_e$ particles produced by $e_+,e_-$ processes. These are small but significant in final impact in standard early universe models. They affect both the "effective number of neutrino species", and the final photon to neutrino temperature ratio.

But before looking at these perturbations, we must correct the MFT calculation of the equilibrium part a little.  We now begin with a 2D system with our two beams at relative angle $\theta$, so that we have doubled the number of beam directions to 4. Then after finishing setting up the equations and initial conditions we can let $\theta \rightarrow 0$ if we wish, to get a strict up and down world. Now there are four beam directions 
and each is to be initially populated with 4 sub-beams with flavor assignments chosen from $\pm 1$ in such a way that the none of the choices for the (now) 16 $\nu$'s in the amplitudes from $\sigma_3^{j}$ replicate the combined flavor and direction of another choice. When $\theta \rightarrow 0$ in this construction
so that two pairs of sub-beams are going in a common direction we could be tempted to say ``O.K.,
life simplifies because if we had an initial beam with the assignments of, say, $[1,-1, 1,-1]$ \newline
(for [$\nu$ flavor, $\nu$ flavor, $\bar\nu$ flavor, $\bar\nu$ flavor]) and one with  $[1,1, -1,-1]$ in the same direction this gives
$2\times [1,0,0,-1]$, and we should then end up with 8 beams rather than 16."

This is not allowed. The reasons give us the right opportunity to discuss general coherence issues at a point at which the reader can see how they matter:
\newline
\newline
A. The initial values of the sub-beams' flavors are prepared by ordinary collisions in the expanding early universe plasma. Ordinary $\nu$ flavor mixing is much slower than our time scale, and furthermore does not supply a seed for a fast process of the type considered here. The coherent particle mixture implicit 
in $2\times [1,0,0,-1]$ is not contained in the initial system. 
\newline
\newline
B. However when we choose the flavor representation there is another kind of coherence, the one that drives fast processes. It comes about because the two-body reactions that drive fast processes are totally forward. Each neutrino with momentum ${\bf p}$ that is included in a beam can have a different phase. But that phase factor is attached to the momentum state and it changes in time not at all for the line ${\bf p}$ (as long as de-phasing effects of $\nu$ mass are neglected.) And it does not enter into the equations in flavor space.
\newline

We now add another pair of beams, numbers 17 and 18, which we refer to collectively as a ``probe", where each contains .001 times the number of $\nu$'s as does each beam the background state. This 
.001 factor does not break the above rule of choosing only [$ \pm 1/,\pm 1 ,\pm 1,\pm 1$] states, since in principle it can be removed from the equations with yet another rescaling.  For the probe we choose initial values of equal numbers of $\nu_e$ ,$\bar \nu_e$, and no  $\nu_x$ ,$\bar \nu_x$. 
Solving ({\ref{array}) now in the $N_a=18$ case with the initial conditions as sketched above, we obtain the data shown in fig. 2.

\begin{figure}[h] 
 \centering
\includegraphics[width=2.5 in]{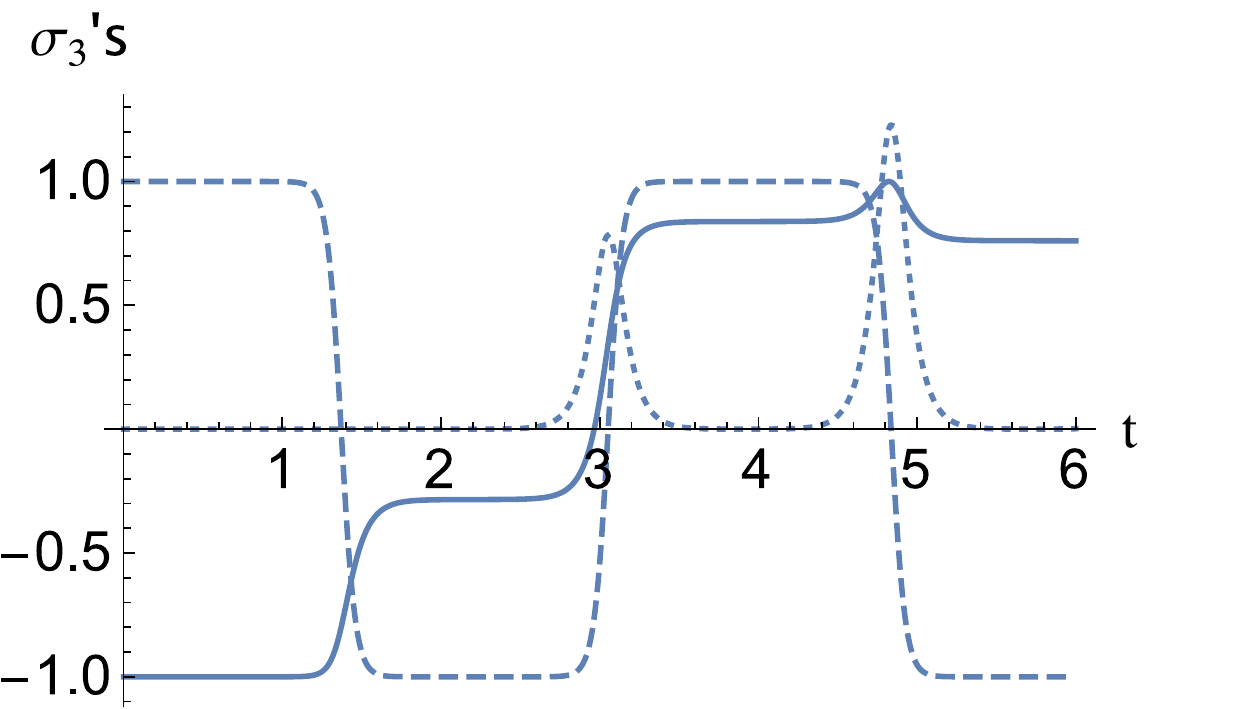}
 \caption{ \small } 
 Solutions of (\ref{array}). $\sigma_3=n_x-n_e$ Dashed curve:  $\sigma_3^1(t)$ in the $\theta\rightarrow 0$ limit. Dotted curve: 
 $ 10^3 (\sigma_3^1(t)  +\sigma_3^2(t))  $.
 Solid curve: $ 10^3\sigma_3^{\rm probe}(t)$. The unit of time is the ``fast scale", $[G_F n_\nu ]^{-1}$.
 \label {fig. 2}
\end{figure}
 
 All of the $\sigma_3^j$, taken one at a time are either the same or the negative of $\sigma_3^1$, as plotted
in the dashed curve, when $\theta \rightarrow 1$, until the second x-axis crossing point of the $\sigma_3^1(t)$, and with fluctuations of order  .1 \% thereafter. The latter appear to be completely due to the perturbation of the main beams by the probe. Thus for at least this long, the flavor equilibrium of the basic 16 beam system has not been disturbed in any way that is obvious.  Beam by beam, though, a lot is going on. 

The point of the above exercise was to look at the ``probe mode"
and to see how it fared. The solid curve shows that it underwent radical changes beginning at the turn-over time
for the individual components of the equilibrium system, continuing in a weird staircase to a complete turn-over of itself.
Thus we conclude that the $N_a=16$ state that was feigning death, from a mean field standpoint, still had a role to play. For brevity, we shall call it the ``possum". Just by being there it was able to induce the 
plotted changes in the probe.
This is of great interest to the standard model as we reach the era of 
${\rm e_+,e_- }$ depletion and the decoupling of the $\nu$ cloud, since the rates are important for the surplus number of 
$\nu_e, \bar\nu_e$ (also carrying higher average energy) to share these properties with the other flavors.

Even in the absence of the probe's perturbation a rate of sterile neutrino production will be affected, for a wide range of sterile $\nu_s$ parameters. The production calculation normally involves the small amplitude oscillatory mixing of an active flavor state with the sterile state of the exactly the same momentum q, with the subsequent freeing of the sterile part in a collision. In our environment of the early universe at temperature of 1-3 MeV, the time between collisions is of order 10$^{10}$ times our turn-over time (which is less than 1 cm./c ), so that net production remains low, due to the requirement of many collisions. 

But now suppose that for a moment all of the sterile-active coupling is to the $\nu_e$ flavor. Then at some moment in time a particular 
 $\nu$ with momentum ${\bf p_1}$ in a state of flavor $\nu_e$ is proceeding along with its small
 sinusoidal (in time) mixing with the $\nu_s$, but then the possum acts to turn the the $\nu_e$ component into a $\nu_x$, leaving the $\nu_s$ part untethered to its parent. The systematic calculation of the influence of the fast processes on the mixing with steriles requires the introduction into ({\ref{array}) of four more amplitudes, and will be deferred until a later publication. But early estimates
 indicate important changes in outcome.

When we start with multiple beams in our calculation 
we need to be certain that their momentum sets are disjoint or nearly so, even when the beams are in the same ``direction".  Note that for the case of a Fermi distribution with zero chemical potential the probability is
$(- \pi^2/9+ \zeta[3])/ \zeta[3]) \approx .087$  independent of temperature,
 that the momentum of a randomly drawn state, matches  \underline{any} of the momenta in another draw, in an isotropic drawing. This gives so much overkill that the enhancements due to the narrow angles of the bundles comprising the system remain inconsequential.

\section{4. Two and three dimensions}

The up-down, 1D, 16 beam framework that we used in the previous section can be used in a planar geometry where there is an angle $\theta$ between 0 and $\pi/2$ between two directions. It only requires appropriate use of a different $g_{j,k}$ that takes
 values [$1,-1,0,(1+\cos \theta)/2,(1-\cos \theta)/2)$] instead of just [$1,-1,0$]. The results for all of the curves plotted
 in fig. 2 are nearly identical in shape for each angle  $\theta$ but with the time scales decreasing in linear fashion, by 
 20\%, as we reach the halfway point $\theta=\pi/4$. We are encouraged by this result, on our path to 3D isotropy.

We could write a Mathematica program for a case with $N_b$ times as many beams at regularly spaced angles, now with $N_a=16 N_b$ but there is no way it could run on our lap-top. We believe that we have 
an inferior approach that works by running a randomized system repeatedly, many, many times. 
It qualitatively confirms the special cases reported here, and extended to multi-angle cases, with rough 3D isotropy, confirms the existence of the effects in this system..

In considering the case of two up-beams (per flavor), in the limit in which the angle between them goes to zero, it might seem that we have needlessly kept the sixteen beam framework and that eight would have sufficed. But the beams should be thought of as very narrow angular cones, in any case, so that the result of the $\theta \rightarrow 0$ limit should apply, rather than a calculation with $\theta = 0$, {\it ab initio}. 

\section{5. Initiation and extinction}

Before reaching firm conclusions much more analysis must be devoted to initiation and extinction.
It may be that after the first sub-system reversal, as we get to the time when the system approaches its completely flipped over initial state ($\zeta=1$ in fig. 1, half way through this plateau) the story ends due to tiny random fluctuations in the environment.  
But surely thinking about incubation and extinction of our modes is easier in this venue than it is in the supernova application. The early universe has the advantage, because of its translational invariance,  strict isotropy and strict balancing of number of $\nu$'s and $\bar\nu$'s.

It might appear that we have assumed a super-long range phase coherence of some kind in order to sustain the long time build-up of these modes, in contradiction to
basic causal principles. To address this concern, in the up-down case, 
 an initial ``up" beam up is fragmented into a long sequence of mini ``up" beam segments, each of length much less than the turn-over distance. We follow the flavor density matrix of one of these segments as it serially encounters a succession of down segments. 
As one of these smaller packages passes through a similar construction of similar guys coming in the other direction, the exact momenta $q_j$ encountered are different in each successive experience; the coherence is in the flavor densities that we assume had evolved according to the same rules in each mini-beam starting at about the same time in the past. At each point in the big plateau of preparation for the turnover, our up-beam segment meets a down segment that we assume was hatched at about the same time

We have checked this picture with a code, where we follow a test mini-beam through a 
whole series of oncoming mini-beams, which originated at about the same time but have passed through entirely different parts of the medium. Time is advanced somewhat jerkily, but carefully following the separate ``up" and ``down" flavor wave-functions' time dependence for each package. The outcome
is essentially the same as in our original picture in which there appeared to be simultaneous interaction 
over the entire distance required for at least one inversion. The initial mini beams can be given arbitrary all-over phases with no change in the results. All of this, it must be said, was calculated with the simplest two flavor model, with the $\sigma^{(e)}_3 \sigma^{(x)}_3 $ coupling omitted so as to produce a two stream instability without requiring antiparticles. 

\section{6.  Discussion}
In this paper we stayed 
within the standard model of weak interactions and worked within the parameter space of the standard early universe. We conclude that when it comes to refining the results of, e.g., \cite {dol} for the small effects of $\nu-\nu$ scattering on the $\nu$ energy spectra and flavor distributions that emerge in the decoupling era , the physics that we have introduced here will be an essential tool. In particular, we conclude that the sharing with other flavors of the excesses of
of $\bar \nu_e, \nu_e$, both in number and in average energy and produced by interactions involving
 e$_+$,e$_-$ , has been greatly accelerated over the standard case.

Two signature features of standard results that should be changed when our approach is fitted into a genuine calculation, are the  $(11/4)^{1/3}$ temperature ratio (photon to neutrino), and the 
``effective number of $\nu$ species =3.04 ". It seems likely that the temperature ratio will be lowered 
and the the effective number of species will be raised.

In the less well-mapped terrain of putting limits on sterile parameters, there appears to be a harder slog ahead. But the subject certainly demands further attention.

I thank Georg Raffelt  for his interest in the work, and for a number of important corrections and comments.

\end{document}